\documentclass{article}
\usepackage{emulateapj,pstricks,apjfonts}

\righthead{SMITH}
\lefthead{NGC 4410 GALAXY GROUP}

\slugcomment{Astrophysical Journal, in press}

\def\sun{_\odot}

\begin{document}

\title{Interstellar Gas
\\in the 
\\NGC 4410 Galaxy Group
}

\author{Beverly J. Smith}
\affil{CASA, University of Colorado, Box 389, Boulder CO 80309;
beverly@casa.colorado.edu\footnote{Now at Physics Department, Box 70652,
East Tennessee State University, Johnson City TN  37614-0652.}}

\begin{abstract}

We present new radio continuum, 21 cm HI, and 2.6 mm CO data for the 
peculiar
radio galaxy NGC 4410A and its companion NGC 4410B, 
and compare with available optical
and X-ray maps.
Our radio continuum maps show an asymmetric double-lobed structure, with
a 
high surface brightness lobe 
extending
3\farcm6 ($\sim$100 kpc)
to the southeast
and a
6\farcm2 ($\sim$180 
kpc)
low surface brightness feature in the northwest.
Molecular gas is abundant in NGC 4410A, with 
M$_{H_2}$ $\sim$ 4 $\times$ 10$^{9}$ M$\sun$ (using
the standard Galactic conversion factor), but is undetected in NGC 4410B.
HI is less abundant,
with
M$_{HI}$ $\sim$ 10$^9$ M$\sun$ for the pair.
Our HI map shows a 3 $\times$ 10$^8$ M$\sun$ HI tail 
extending 1\farcm7 (50 kpc) to the southeast of the pair, 
coincident with a faint optical tail and
partially
overlapping with the southeastern
radio lobe.
The HI tail is anti-coincident with a 2$'$ (56 kpc) long
X-ray structure aligned with
a stellar bridge that 
connects the pair to a third galaxy.
If this X-ray emission is associated with the group,
we infer 3 $-$ 8 $\times$ 10$^8$ M$\sun$
of hot gas in this feature.  This
may be either intracluster gas or shocked gas associated with
the bridge.

Our detection of abundant interstellar gas in
this pair, including an
HI-rich tidal tail near the southeastern radio lobe,
suggests 
that the observed distortions in this lobe
may have been caused by the interstellar medium in this system.
The gravitational interaction of the two galaxies and
the subsequent motion of the interstellar medium
in the system
relative to the jet may have produced sufficient
ram pressure to bend and distort the radio jet.
An alternative hypothesis is that the jet was distorted
by ram pressure due to an intracluster medium, although the 
small radial velocity of NGC 4410A relative to the group
and the 
lack
of diffuse X-ray emission in the group makes this
less likely unless the group is not virialized or
is in the process of merging with another group.

Using our VLA data,
we also searched for HI counterparts
to the other 
ten known
members of the NGC 4410 group and CO from three other galaxies in
the inner group.
In our velocity range of 6690 km s$^{-1}$ $-$ 7850 km s$^{-1}$, we detected
six other
galaxies above our HI sensitivity limits of 2 $\times$ 10$^8$ M$\sun$ for the
inner group and 4 $\times$ 10$^8$ M$\sun$ for the outer group.
The total HI in the group is 1.4 $\times$ 10$^{10}$ M$\sun$, 80$\%$ of which
arises from four galaxies in the outer group.
Three of these galaxies (VCC 822, VCC 831,
and VCC 847) are spirals with M$_{HI}$/L$_B$ ratios
typical of field galaxies,
while FGC 170A
appears to be a gas-rich dwarf galaxy
(M$_B$ $\sim$ $-$18; M$_{HI}$ $\sim$ 3 $\times$ 10$^9$ M$\sun$).
In the inner group, 
the SBa galaxy
NGC 4410D (VCC 934)
was detected in HI and CO
(M$_{HI}$ $\sim$ 5 $\times$ 10$^8$ M$\sun$ and
M$_{H_2}$ $\sim$ 
8 $\times$ 10$^8$ M$\sun$)
and has a 
1$'$ (28 kpc) long HI tail that points towards the nearby disk galaxy NGC 4410F.
NGC 4410F was also detected in HI
(M$_{HI}$ $\sim$ 4 $\times$ 10$^8$ M$\sun$).
The galaxies in the inner group appear to be somewhat deficient
in HI compared to their blue luminosities,
suggesting phase changes driven by galaxy-galaxy or galaxy-intracluster
medium encounters.
 
\end{abstract}


\section{INTRODUCTION}

\subsection{The Interstellar Medium in Radio Galaxies}

There is increasing evidence that the large radio structures
found near luminous radio galaxies both affect and are affected by
their environment.  They
may be distorted 
by motion through an intracluster medium (\markcite{m72}Miley et al. 1972;
O'Dea $\&$ Owen 1985),
or by
an encounter with the interstellar medium in the host
galaxy 
(e.g., \markcite{dY81}de Young 1981; \markcite{vB84}van Breugel et al. 
1984, \markcite{vB85a}1985a),
a companion
galaxy (\markcite{sbc85}Stocke, Burns, $\&$ Christiansen 1985; 
\markcite{bc93}Borne 
$\&$ Colina 1993;
\markcite{hll98}Harris,
Leighly, $\&$ Leahy 1998), or a galaxy in the process of merging with
the host galaxy (\markcite{vB86}van Breugel et al. 1986).
A gravitational encounter between two galaxies
or clusters of galaxies may disrupt previously-existing
radio lobes by perturbing the interstellar or
intracluster medium (\markcite{co91}Comins $\&$ Owen 1991;
\markcite{bc93}Borne 
$\&$ Colina 1993).
In some systems, an impact between a jet and the interstellar medium appears to have 
triggered star formation 
(\markcite{dY81}de Young 1981; 
\markcite{vB85b}van Breugel et al. 1985b; 
\markcite{vBd93}van Bruegel $\&$ Dey 1993;
\markcite{g98}Graham 1998).

To investigate the interplay between the radio lobes and the interstellar medium
in radio galaxies, 
detailed studies of the interstellar matter in
nearby radio galaxies are essential.
Although there have been a number of narrowband optical imaging surveys of 
radio galaxies (e.g., \markcite{hnj87}Hansen, Norggard-Nielsen, $\&$ Jorgensen 1987;
\markcite{b88}Baum et al. 1988;
\markcite{mut92}Morganti, Ulrich, $\&$ Tadhunter 1992;
\markcite{m95}McCarthy et al. 1995; \markcite{hbf96}Hes, Barthel, $\&$ Fosbury 1996), 
measuring the distribution of ionized gas in more than 100 radio galaxies,
studies of the neutral gas in radio galaxies have been scarce.
A few pioneering single-dish searches for the 21 cm HI line 
(\markcite{dbo82}Dressel, Bania, $\&$ O'Connell 1982;
\markcite{bb83}Bieging $\&$ Biermann 1983;
\markcite{j83}Jenkins 1983) 
and the 2.6 mm CO line 
(\markcite{m89}Mirabel et al. 1989;
\markcite{m93}Mazzarella et al. 1993)
in radio galaxies have detected only a handful of radio galaxies.
There have been a number of high resolution interferometric
studies of HI absorption towards the nuclei of 
radio galaxies (e.g., \markcite{vG89}van Gorkom
et al. 1989), however, the 
large-scale distribution of HI
emission has been 
successfully mapped
in only a handful of radio galaxies, including
Centaurus
A (\markcite{vG90}van Gorkom et al. 1990; \markcite{s94}Schiminovich et al. 1994), 
NGC 1052 (\markcite{vG86}van
Gorkom et al. 1986), NGC 4278 (\markcite{r81}Raimon et al. 1981; \markcite{l92}Lees 1992),
and PKS B1718-649 (\markcite{vc95}V\'eron-Cetty et al. 1995).

In this paper, we present a detailed study of the interstellar matter
in the nearby low luminosity radio galaxy NGC 4410A and its companions,
and compare with other radio galaxies.
Throughout this paper we use a Hubble constant of 
H$_{\rm o}$ = 75 km s$^{-1}$ Mpc$^{-1}$.

\subsection{NGC 4410} 

Optical images of NGC 4410A, classified as an Sab ? Pec galaxy by
\markcite{rc3}de Vaucouleurs et al. (1991), reveal a prominent bulge
surrounded by an extended ring or loop-like structure
(\markcite{h86}Hummel et al. 1986).
This structure is clearly visible in 
Figure 1a, 
a broadband red Hubble Space Telescope (HST) archival image\footnote{
Based on observations made with
the NASA/ESA Hubble Space Telescope, obtained from the data archive
at the Space Telescope Science Institute.  STScI is operated by
the Association of Universities for Research in Astronomy, Inc. under
NASA contract NAS 5-26555.}
of NGC 4410A.
In this image,
a number of bright knots are apparent in the eastern half
of the ring, and an especially luminous knot is visible
southeast of the nucleus.
Narrowband H$\alpha$+[N~II] imaging and optical spectroscopy
show that these knots are extremely luminous
H~II region complexes with
L$_{H\alpha}$ $\sim$ 10$^{40}$ $-$ 10$^{41}$ erg s$^{-1}$
(\markcite{d99}Donahue et al. 2000).
The western portion of this ring contains filamentary ionized
gas which appears to be shock-ionized (\markcite{d99}Donahue
et al. 2000).
The presence of interstellar dust is apparent in the HST image;
beautiful filamentary dust lanes
are visible southeast of the nucleus and along the eastern half
of the ring, extending into the center of the galaxy.

NGC 4410A is strongly interacting with the nearby galaxy NGC 4410B
(Figure 1b; Donahue et al. 2000), which 
has been classified as S0~? Pec (\markcite{rc3}de
Vaucouleurs et al. 1991) or E (\markcite{h86}Hummel, Kotanyi,
$\&$ van Gorkom 1986).
These two galaxies are part of a sparse group containing
at least a dozen galaxies which lies behind
the Virgo cluster (Table 1a).
The inner 10$'$ diameter of this group contains five of these galaxies,
at least four of which are strongly interacting (Figure 1c;
Donahue et al. 2000).
NGC 4410A+B may be connected by a stellar bridge to the S0 galaxy NGC 4410C (IC 
790), which in turn is connected to the SBa galaxy NGC 4410D (VCC 934). 
In addition to the possible bridge to NGC 4410C,
NGC 4410A+B has a long (2$'$ $\sim$ 55 kpc) tail to
the southeast and another extension to the northwest
(\markcite{h86}Hummel et al. 1986;
Figure 1c).
A larger field of view (30$'$ diameter) optical picture of the NGC 4410 group
is presented in Figure 1d, where the
Digitized Sky Survey\footnote{The Digitized
Sky Survey was produced at the
Space Telescope Science
Institute under U.S. Government
grant NAG W-2166.
This image is a 
digitized version of 
a photographic plate from the 
Second Palomar Observatory
Sky Survey (POSS-II) made by
the California Institute
of Technology with funds from
the National Science Foundation,
the National Geographic Society, the Sloan
Foundation, the Samuel Oshin Foundation, and the Eastman
Kodak Corporation.
} 
(DSS) optical image is shown.
Table 1a contains the twelve known galaxies in the inner 30$'$
diameter of the NGC 4410 group,
Table 1b gives foreground and background galaxies in the NGC 4410
field, and Table 1c lists additional galaxies in the
field with unknown redshifts.
These 27 galaxies
are identified in Figure 1d.

Both NGC 4410A and NGC 4410B have been classified
as `Low Ionization Nuclear Emission Region' (LINER) galaxies
based on their optical spectra (\markcite{mb93}Mazzarella $\&$
Boroson 1993; \markcite{d99}Donahue et al. 2000).
NGC 4410A+B has a 4.8 GHz luminosity of 1.5 $\times$ 10$^{23}$ W Hz$^{-1}$
(\markcite{cdb91}Condon, Frayer, $\&$ Broderick 1991).  This luminosity
is near the dividing line between starbursts and radio-loud
active galaxies.
Galaxies in which the radio continuum is powered by
star formation, as indicated by a high infrared to radio 
continuum ratio, have L$_{4.8~GHz}$ $\le$ 10$^{23}$ W Hz$^{-1}$
and tend to be spirals, while systems with
higher radio luminosities have lower FIR/radio ratios, indicating
radio-loud active nuclei, and tend to be classified as elliptical or S0
(\markcite{c91}Condon et al. 1991).
The 
NGC 4410A+B pair has a far-infrared luminosity (as in \markcite{l85}Lonsdale
et al. 1985) of 3.9 $\times$ 10$^9$ L$\sun$
(\markcite{mbb91}Mazzarella, Bothun, $\&$ Boroson 1991), giving
a far-infrared to 4.8 GHz luminosity ratio 
(as in Condon et al. 1991)
of $\sim$2.8,
two orders of magnitude less
than that expected for starbursts (Condon et al. 1991), 
placing it firmly
in the radio galaxy class. 

Radio continuum maps 
of NGC 4410A+B show that most of the radio emission
originates from
a very large
(4\farcm5 $\times$ 3$'$; 120 kpc
$\times$ 80 kpc) 
knotty structure extending
to the southeast of NGC 4410A+B
(\markcite{h86}Hummel et al. 1986; \markcite{b92}Batuski et al. 1992).
A compact radio continuum source coincident with the optical
nucleus of NGC 4410A is
also observed, indicating that NGC 4410A is the source of the
radio emission.
This peculiar radio structure has been
attributed to the motion of a low luminosity 
radio galaxy through intracluster gas
(\markcite{sb87}Stocke $\&$ Burns 1987),
or 
to a low luminosity radio galaxy disturbed 
by an interaction (\markcite{h86}Hummel
et al. 1986).

The NGC 4410 group was observed at energies between 0.1 and 2.4 keV
by the ROSAT X-ray satellite, using both the High Resolution
Imager (HRI) and Position Sensitive Proportional Counter (PSPC) 
(\markcite{thj99}Tsch\"oke, Hensler, $\&$ Junkes 1999).
The HRI map shows a point-like source associated with NGC 4410A and
a diffuse halo extending 10$''$
to the southeast, while
the low resolution PSPC map shows some faint diffuse emission
extending eastward from the NGC 4410A+B pair.
The PSPC X-ray spectrum
of NGC 4410A+B can be deconvolved into two components,
a power-law spectrum and a Raymond-Smith spectrum,
presumably arising from the compact and extended sources, respectively.

To better understand the relationship
between the interaction, the radio
emission, and the interstellar medium in the NGC 4410 group,
we have obtained new radio continuum, 21 cm HI, and CO (1 $-$ 0) observations 
of this group, and have compared these with available
optical and X-ray data.
In Section 2 of this paper, we describe the observations
and data reduction.  In Section 3, we summarize
the results, which are discussed in detail
in Section 4.
Conclusions are given in Section 5.

\section{OBSERVATIONS AND DATA REDUCTION}

\subsection{The 21 cm HI Observations}

The NGC 4410 group was observed in the 21 cm HI line
using the D configuration of the NRAO\footnote{
The National Radio Astronomy Observatory
is a facility of the National Science Foundation, operated
under cooperative agreement by Associated Universities,
Inc.
} 
Very Large Array
on 1997 December 22.
The two IF mode was used with 31 195 kHz channels and on-line Hanning smoothing,
giving
a spectral resolution of 
43 km s$^{-1}$ and 
a total
bandpass of 1330 km s$^{-1}$, centered on a velocity of 7252 km s$^{-1}$.
The total on-galaxy observing time was 8 hours.
During this observing run, 3C 286 was observed
for 20 minutes for flux and bandpass calibration,
and a nearby phase calibrator was observed every 30 minutes
for 5 minutes.

The data were calibrated and maps were made
using the NRAO Astronomical Image Processing
System (AIPS).  
After calibration, the AIPS task UVLIN was utilized to 
remove the continuum,
and CLEANed line-only datacubes were produced
using the AIPS routine IMAGR.
Inspection of these channel maps 
shows that 
eight galaxies in the field were detected in HI
(see Table 2) and
line 
emission is present in 22 
channels, from 6777 $-$ 7685 km s$^{-1}$
(see Sections 3.2 and 3.3).
To best isolate the HI emission from the different
galaxies in the field,
the continuum subtraction
process was done four times,
using four different sets of continuum channels.
These four sets of maps were then combined into
a single cube, selecting the appropriate region around
each galaxy.
For highest sensitivity, a set of maps was made with natural
weighting,
giving a beamsize of 59\farcs38
$\times$ 54\farcs67 and an rms of 0.21 mJy beam$^{-1}$
channel$^{-1}$.
Higher resolution maps were also made 
using 
the \markcite{95}Briggs
(1995) robustness weighting scheme with different robustness parameters.
These additional
maps did not show significantly different morphologies, and so are not
discussed further here.

A final continuum map was made by combining the line-free
channels at the ends of the bandpass.
The final CLEANed, primary beam-corrected,
naturally-weighted continuum map of the NGC 4410 field is superposed
on the optical DSS image
in Figure 2.
The noise level in this map is 0.17 mJy beam$^{-1}$ (4.6 $\times$ 10$^{-5}$
mJy arcsec$^{-2}$), compared to 0.13 mJy beam$^{-1}$ (3.1 $\times$ 10$^{-4}$
mJy arcsec$^{-2}$) in the 
higher resolution
\markcite{h86}Hummel 
et al. (1986) C Array map.
The inner portion of the channel maps, showing the central
five galaxies of the NGC 4410 group, is presented in Figure 3.

We created moment maps using the AIPS routine MOMNT.
For these maps,
a scratch copy of the data cube was convolved 
with a 60$''$ Gaussian.  Only pixels above 0.4 mJy beam$^{-1}$
in this smoothed cube were including in deriving 
an HI intensity
map from the original unsmoothed data cube.
To construct an HI velocity map,
a flux cutoff of 0.7 mJy beam$^{-1}$ was used.
These moment
maps were then corrected for primary beam attenuation.

The naturally-weighted HI intensity map of the inner
portion of the NGC 4410 field is
displayed in Figure 4a, superposed on the 
Donahue et al. (2000)
R band image.  For 
comparison, the radio continuum map 
and the low resolution ROSAT PSPC X-ray map from 
\markcite{thj99}Tsch\"oke et al. (1999) 
are also overlaid on the optical map in Figure 4b, along with
the HI map.
In Figure 5, a larger field of view HI intensity map 
is overlaid on the optical DSS image, and
the global HI spectra for the detected galaxies are plotted
in Figure 6.

In Table 2, we provide the HI line fluxes
for the detected galaxies in the group, their central velocities,
and their line widths, along with upper limits for
the other group galaxies in the observed velocity range.
The HI masses for the detected galaxies 
range from 4 $\times$ 10$^8$ M$\sun$ 
$-$ 
10$^9$ M$\sun$ for 
the galaxies in the inner core of the group, to 
2 $-$ 3 $\times$ 10$^9$ M$\sun$ for the four outer galaxies.
These masses are provided in Table 4,
along with the blue luminosities
and M$_{HI}$/L$_B$ ratios.

\subsection{The CO (1 $-$ 0) Observations}

Six positions in the 
NGC 4410 system were observed in
the 2.6mm 
CO (1 $-$ 0) line during
1997 April and May, 1998 June, and 1999 January, 
February, and November,
using the 3mm SIS
receiver on the NRAO 12m telescope.
These six positions 
are the centers of the five innermost galaxies in the
group (NGC 4410A, B, C, D, and F) 
and a position in the southeastern HI tail of
NGC 4410A+B (Table 3).
Two 256$\times$2 MHz filterbanks, one for each
polarization, were used for these observations,
providing a total bandpass of 1330 km s$^{-1}$ with
a spectral resolution of 5.2 km s$^{-1}$.
The positions and central velocities used 
for these observations
are those tabulated in Table 1.
A nutating subreflector with a beam throw of 3$'$
in the azimuthal direction
(1$'$ for NGC 4410F)
was used for these observations, taking care to avoid
chopping on other galaxies, tails, or bridges in the group.
Each scan was 6 minutes
long. The beamsize FWHM is 55$''$ at 115 GHz.
The pointing was checked 
hourly using 3C 273 and was consistent
to 10$''$.  The system temperatures ranged
from 300 to 400~K.  
Calibration was accomplished using an ambient temperature chopper wheel.

The CO spectra for all six observed positions 
are displayed in Figure 7.
CO was detected in NGC 4410A and NGC 4410D,
but was not found towards NGC 4410B, NGC 4410C, NGC 4410F, or
the NGC 4410A+B tail.
The CO line fluxes are given in Table 3,
along with the central velocities 
of the lines, the line widths, and the rms noise levels.
In Table 4,
the molecular gas masses and upper limits for the observed
galaxies in the group are given, calculated assuming the standard
Galactic I$_{CO}$/N$_{H_2}$ ratio 
(\markcite{b86}Bloemen et al. 1986).

\section{RESULTS}

\subsection{NGC 4410A+B}

\subsubsection{The Radio Continuum Map}

As noted earlier,
the radio continuum morphology of NGC 4410A+B is highly peculiar.
The VLA C Array
observations 
of
Hummel et al. (1986) showed a knotty distorted 
structure
extending
4\farcm5 (120 kpc)
to the southeast of the optical galaxies.
This feature is also visible in our new D Array data (Figure 2 and Figure 4b).
In addition, our data has revealed the existence of another
much lower surface
brightness feature, extending 6\farcm2 (180 kpc) to the northwest
of NGC 4410A (Figure 2).
Like the southern lobe, the northern feature is distorted.
The ridge that connects this lobe to the southern lobe is offset to
the east with respect to NGC 4410A.

Our data gives
a total 20 cm continuum flux from NGC 4410A+B of 403 $\pm$ 11 mJy, with
23 $\pm$ 7 mJy of this coming from the northern lobe.
For comparison,
\markcite{k80}Kotanyi (1980) and \markcite{h86}Hummel 
et al. (1986) found 20 cm flux densities for NGC 4410A+B of 
392 $\pm$ 21 mJy (as corrected by
\markcite{h86}Hummel et al. 1986) and 340 $\pm$ 30 mJy,
respectively.

\subsubsection{The 21 cm HI Data}

The distribution of interstellar gas in NGC 4410A+B is 
strongly disturbed (Figure 4a).
The peak of the HI distribution lies $\sim$25$''$
to the southeast
of NGC 4410A, in the direction of the 
prominent H~II region visible in the Hubble Space
Telescope image (Figure 1a).
Extending to the southeast of
this peak is a tail 
1\farcm7 (50 kpc) long,
containing M$_{HI}$ $\sim$ 3 $\times$ 10$^8$ M$\sun$.
This gaseous structure is associated
with a faint optical tail (Figure 4a).
A second structure containing M$_{HI}$ $\sim$ 10$^8$ M$\sun$
extends $\sim$0\farcm5 (14 kpc)
to the east of the optical galaxies,
near the stellar bridge connecting the pair with NGC 4410C.
A third extension containing M$_{HI}$ $\sim$ 10$^8$ M$\sun$ 
lies $\sim$1$'$ (28 kpc) due south of the
pair.
The total HI line flux for NGC 4410A+B is 0.55 $\pm$ 0.15 Jy km s$^{-1}$,
corresponding to 1.3 $\times$ 10$^{9}$ M$\sun$ of atomic
hydrogen.  
This measurement is 
consistent with previously-published single-dish fluxes of 
$\le$0.51 Jy km
s$^{-1}$ 
(\markcite{bb83}Bieging $\&$ Biermann 1983) and 
0.74 Jy km s$^{-1}$ (\markcite{h89}Hoffman et 
al. 1989), both made 
with 
the Arecibo 3\farcm2 beam.

Although both the radio continuum and HI line
emission of NGC 4410A+B extend to the southeast of NGC 4410A+B,
their detailed distributions are different (Figure 4b).
The HI is distributed
in a narrow tail-like structure extending southwest, with two shorter 
extensions to the south and east.
The continuum is more widely distributed than the HI gas,
extending further in the east-west
direction,
although it has a similar north-south size.

The HI spectrum for NGC 4410A+B (Figure 6)
shows a peak at $\sim$7350 km s$^{-1}$,
close to the optical velocity of NGC 4410A (7440 km s$^{-1}$;
Table 1).  NGC 4410A+B also has 
a second weaker HI peak at $\sim$7140 km s$^{-1}$,
near the optical velocity of the luminous HII region
to the southeast of NGC 4410A and the H~II regions
in the ring (\markcite{d99}Donahue et al. 2000).
The VLA line width and the observed velocity range for NGC 4410A+B 
are consistent with those of the single-dish spectrum of
\markcite{h89}Hoffman et al. (1989).

The HI velocity field of NGC 4410A+B
is very disturbed.
Because of these
peculiarities, the velocity
structure is better 
seen in the channel maps (Figure 3) than in standard
intensity-weighted
first moment maps.
In Figure 3, 
the NGC 4410A+B HI tails are visible as 3 $-$ 5$\sigma$ structures 
between 
7295 km~s$^{-1}$ and
7382 km~s$^{-1}$.
The southeastern tail is somewhat blueshifted 
(7295 $-$ 7338 km~s$^{-1}$)
with respect to the feature extending due south (7382 km s$^{-1}$).
Another 
5$\sigma$ clump at the end of southeastern tail is 
blueshifted 
130 km s$^{-1}$ 
from the rest of the tail 
to 7166 km s$^{-1}$.
The central HI component
is visible from
7122 $-$ 7382 km s$^{-1}$, from the optical velocity of
the NGC 4410A nucleus (7440 km s$^{-1}$) to the velocities of 
the H~II regions southeast of the nucleus and in the ring 
(7140 $-$ 7300 km s$^{-1}$; \markcite{d99}Donahue et al. 2000).

\subsubsection{The Molecular Gas}

The CO line was detected towards
NGC 4410A but not towards NGC 4410B or 
the observed position in the southeastern HI tail (Figure 7).
Thus the molecular gas in this pair is concentrated in NGC 4410A.
The CO line of NGC 4410A is very broad, covering
a total velocity range of $\sim$750 km s$^{-1}$, from 7000 km s$^{-1}$
to 7750 km s$^{-1}$.
This spectum has three peaks, at $\sim$7140, 7300, and 7565 
km s$^{-1}$.
The first two velocity components have HI counterparts (Figure 6a);
the high velocity feature at 7565 km s$^{-1}$ does not.  
The ionized gas, however, covers a velocity range similar to
that of the CO (\markcite{d99}Donahue et al. 2000).
In the eastern portion of the NGC 4410 ring,
the H~II regions have optical velocities between 7140 km s$^{-1}$
and 7300 km s$^{-1}$, while the shock-ionized gas
in the western part of the ring has an optical velocity
of 7500 km s$^{-1}$ (\markcite{d99}Donahue et al. 2000).
The optical ring lies entirely in the CO beam, therefore, it is
possible
that the low velocity CO gas lies in the eastern portion
of the ring and is associated with the star formation regions,
while the high velocity molecular gas is associated with the 
western filamentary feature.
However, the nucleus of NGC 4410A has broad optical lines (FWHM $\sim$ 600 
km s$^{-1}$; Donahue et al. 2000),
so we cannot rule out a more concentrated location
for the molecular gas.
NGC 4410A+B has a total H$_2$ mass of 3.9 $\times$ 10$^{9}$ M$\sun$;
about 40$\%$ of this is in the high velocity component at 7565 km s$^{-1}$.

\subsubsection{The X-Ray Distribution}

The difference between the HI and X-ray morphologies of
NGC 4410A+B is remarkable (Figure 4b).
In the X-ray map, 
a faint elongation
extending 2$'$ (56 kpc) to the east 
of the NGC 4410A nucleus
is apparent.
This lies along the 
stellar bridge connecting NGC 4410A+B with NGC 4410C.
In contrast, the extended HI is found mainly along the 
southeastern tail of NGC 4410A+B rather than the bridge, although
there is a hint of a HI extension near the base of the X-ray extension,
near the optical bridge.

The X-ray feature is detected at a 3$-$8$\sigma$ level above
the X-ray background, at 4 $-$ 10 $\times$ 10$^{-7}$
counts s$^{-1}$ arcsec$^{-2}$.
Assuming a 1 keV Raymond-Smith plasma and the 
Galactic N$_H$ value of 1.7 $\times$ 10$^{20}$ cm$^{-2}$
(\markcite{dl90}Dickey $\&$ Lockman 1990),
we find an 
unabsorbed 0.1 $-$ 2.4 keV 
surface brightness in the bridge of
5 $-$ 13 $\times$ 10$^{-18}$ erg s$^{-1}$ cm$^{-2}$ arcsec$^{-2}$,
using the on-line PIMMS conversion tool (Mukai 1993).
If this extended emission is indeed associated with the group (see
Section 4.1), it has
an X-ray luminosity
of 4 $\times$ 10$^{40}$ erg s$^{-1}$, $\sim$10$\%$
of the total X-ray luminosity of NGC 4410A+B.

\subsection{NGC 4410D}

The barred galaxy NGC 4410D was also detected in 
HI and CO, with 
M$_{HI}$ $\sim$ 5 $\times$ 10$^8$ M$\sun$ and
M$_{H_2}$ $\sim$ 8 $\times$ 10$^8$ M$\sun$.
Like NGC 4410A+B, NGC 4410D
has a peculiar HI morphology and velocity structure.
There is a long HI tail 
extending 1$'$ (28 kpc) south from
NGC 4410D towards NGC 4410F, aligned with a faint stellar feature
(Figure 4a).
This structure contains M$_{HI}$ $\sim$ 2 $\times$ 10$^8$ M$\sun$.
There is also a second concentration of HI associated with the
main disk of this galaxy, with a slight 
($\sim$0\farcm5 $\sim$ 14 kpc) 
extension 
northwards from
the eastern edge of NGC 4410D.  

The HI spectrum of NGC 4410D shows two distinct features, both
with relatively narrow (FWHM $\sim$
90 km s$^{-1}$) line widths: one near the optical velocity 
at 6960 km s$^{-1}$
and a second at 7380 km s$^{-1}$ (Figure 6a).
These two velocity components are spatially
separated: the 
component at 7382 $-$ 7425 km s$^{-1}$
is associated with the southern tail, while the
gas at 6907 $-$ 6993 km s$^{-1}$
is associated with the optical disk and northern extension.
The CO line of NGC 4410D (Figure 7) has
a peak velocity of 6960 km s$^{-1}$, consistent
with the low velocity HI peak and the optical
value.

\subsection{NGC 4410F}

The small galaxy NGC 4410F was detected in HI but not
in CO, with M$_{HI}$ $\sim$ 4 $\times$ 10$^8$ M$\sun$
and M$_{H_2}$ $\le$ 5 $\times$ 10$^8$ M$\sun$.
The HI distribution of NGC 4410F appears somewhat disturbed,
being slightly elongated to the south (Figure 4a).
NGC 4410F has such a narrow HI line (Figure 6a) that it is 
almost unresolved at our velocity resolution; it is detected
above 3$\sigma$ in only two channels (Figure 3).

\subsection{The Outer Galaxies in the NGC 4410 Group}

In addition to NGC 4410A+B, NGC 4410D, and NGC 4410F, four
galaxies in the outskirts of the NGC 4410 group were 
detected in HI:  
VCC 822,
VCC 831, VCC 847, and FGC 170A (Figure 5).
These galaxies
have more regular HI distributions and velocity fields than
the galaxies in the inner group.
The intensity-weighted HI velocity maps for these galaxies
are shown in Figures 8a-d, superposed on the optical DSS images.
These maps
show evidence of rotation and 
lines of nodes consistent with
the optical position angles.

These four galaxies
have higher M$_{HI}$/L$_B$ ratios than the galaxies in the inner
group (Table 4).
VCC 822,
VCC 831, and VCC 847 have M$_{HI}$/L$_B$ $\sim$ 0.12 $-$ 
0.26~M$\sun$/L$\sun$, 
typical of field spirals 
(\markcite{hg84}Haynes $\&$ Giovanelli 1984).
FGC 170A has the highest M$_{HI}$/L$_B$ ratio
of the group, 1.2 M$\sun$/L$\sun$, and the lowest blue luminosity,
2 $\times$ 10$^9$ L$\sun$.
This corresponds to an 
absolute B magnitude of $\sim$$-$18, similar to
that of the Large Magellanic Cloud.

\section{DISCUSSION}

\subsection{Comparison with Other Radio Galaxies}

Our detection of a second large radio continuum structure near
NGC 4410 confirms that this galaxy is indeed a double-lobed
radio galaxy, albeit with a distorted radio morphology.
The peculiar ring-like optical structure of NGC 4410A, along with
the existence of the optical and/or HI tails/bridges
extending from the NGC 4410A+B pair, indicates that 
these galaxies have been gravitationally disturbed by a collision
or close encounter.
The hosts of radio galaxies are 
generally elliptical or elliptical-like
(e.g., \markcite{lp87}Lilly $\&$ Prestage 1987; 
\markcite{ow91}Owen $\&$ White 1991).  A number of recent studies 
have claimed a higher rate of morphological peculiarities
in radio galaxy hosts compared to
radio-quiet ellipticals 
(\markcite{heck86}Heckman et al. 1986;
\markcite{cdJ95}Colina $\&$ de Juan 1995; 
\markcite{z96}Zirbel 1996), although this conclusion is controversial
(\markcite{lo95}Ledlow $\&$ Owen 1995).
The peculiar optical morphology of NGC 4410A is
therefore
unusual for a radio galaxy but perhaps not unique.

The HI morphologies and velocity structures
of the galaxies in the inner core of the NGC 4410 group
are quite peculiar compared with normal disk galaxies.
At least two HI tails are detected, 
both with optical counterparts. 
The presence of these stellar counterparts indicates that the HI distortions
are not due to ram pressure stripping by intracluster gas;
tidal interactions between the galaxies are probably predominantly responsible
for these structures.
The five galaxies in the inner core are quite
deficient in HI 
relative to their blue luminosities,
compared with field spirals (e.g., 
\markcite{hg84}Haynes $\&$ Giovanelli 1984).
The tail-like features in this group are not particularly rich
in HI 
(M$_{HI}$ $\sim$ 1 $-$ 4 $\times$ 10$^8$ M$\sun$) compared
with other tidal tails, 
which sometimes have M$_{HI}$ $\sim$ 10$^9$
M$\sun$ (e.g., \markcite{s91}Smith 1991; \markcite{hg96}Hibbard $\&$ 
van Gorkom 1996; \markcite{d97}Duc et al. 1997).
The southeastern tail of NGC 4410A+B has a blue magnitude 
of $\sim$2 $\times$ 10$^9$
L$\sun$ (\markcite{d99}Donahue et al. 2000), giving
M$_{HI}$/L$_B$ $\sim$ 0.2 M$\sun$/L$\sun$ for this feature,
similar to the tails studied by \markcite{s91}Smith (1991)
and \markcite{hg96}Hibbard $\&$ van Gorkom (1996).

The molecular gas mass
of NGC 4410A (4 $\times$ 10$^9$ M$\sun$)
and its broad (FWHM $\sim$
600 
km~s$^{-1}$)
CO line are not unusual
for a radio galaxy.
Out of the ten radio galaxies surveyed by \markcite{m89}Mirabel et al. (1989)
and 
\markcite{m93}Mazzarella et al. (1993), seven have 
CO luminosities similar to that of NGC 4410A, some with similarly broad lines.
In contrast to the 
radio galaxies in these two surveys, 
however,
NGC 4410A+B has a low L$_{FIR}$/M$_{H_2}$ ratio of $\sim$1 L$\sun$/M$\sun$,
indicating a lower rate of star formation compared with the available
molecular gas.

It is not clear at present whether
the amount of HI in NGC 4410A+B, 10$^9$ M$\sun$,
is high 
compared with other radio galaxies, or if the presence of the HI tail
is unusual in a radio galaxy.
The few single-dish HI emission-line surveys of radio galaxies done to date 
(\markcite{dbo82}Dressel et al. 1982;
\markcite{bb83}Bieging $\&$ Biermann 1983;
\markcite{j83}Jenkins 1983) 
have not been sensitive enough to provide strong constraints on the HI masses
of radio galaxies.
Although a number of high spatial resolution HI absorption studies
of radio galaxies have been done (e.g.,
\markcite{vG89}van Gorkom et al. 1989),
only a handful of HI emission line
maps of radio galaxies have been published to date.
Centaurus A has a warped HI disk (\markcite{vG90}van Gorkom et al.
1990) and HI associated with `shells' 
(\markcite{s94}Schiminovich et al. 1994),
NGC 4278 has a large rotating disk of HI
(\markcite{r81}Raimond et al. 1981; 
\markcite{l92}Lees 1992),
NGC 1052 and PKS B1718-649 have extended HI disks
with tidal features
(\markcite{vG86}van
Gorkom et al. 1986; 
\markcite{vc95}V\'eron-Cetty et al. 1995),
while Pictor A, PKS 0349-27, and PKS 0634-20 
were undetected in HI with the VLA
(\markcite{sc96}Simkin 
$\&$ Callcut
1996). 

The X-ray morphology of NGC 4410A+B is rather unusual.
The only other galaxy with a possible
X-ray counterpart to
a tidal feature is the radio galaxy Fornax A
(\markcite{mf98}Mackie $\&$ Fabbiano 1998).
The tails of the eight non-radio galaxy mergers studied
by
\markcite{h94}Hibbard
et al. (1994),
\markcite{rpw95}Read,
Ponman, $\&$ Wolstencroft (1995) and \markcite{rp98}Read $\&$ Ponman (1998)
were not detected by ROSAT.
In both the Fornax A feature and the NGC 4410 bridge, 
the X-rays 
appear to be due to hot gas rather than X-ray binaries
associated with the stellar population in the tidal feature.
The blue luminosity of the bridge between NGC 4410A+B and NGC 4410C
is 2 $\times$ 10$^9$ L$\sun$
(\markcite{d99}Donahue
et al. 2000), 
giving L$_{X}$/L$_{B}$ $\sim$ 2 $\times$ 10$^{31}$ erg s$^{-1}$/L$\sun$.
This ratio is $\sim$75 times higher than that found for normal 
spiral galaxies (\markcite{fkt92}Fabbiano, Kim, $\&$ Trinchieri 1992)
indicating that this X-ray emission is not due to X-ray binaries.

The possibility that the X-ray `tail' near NGC 4410 
is due to hot gas outflowing from the 
luminous star formation regions and/or the active nucleus,
coincidently superposed on an unrelated stellar bridge, also appears to
be ruled out by
the large X-ray luminosity 
and size 
of this feature.
A single-sided X-ray outflow
with similar luminosity and size 
has been found
near the `ultraluminous' galaxy Arp 220
(\markcite{h96}Heckman et al. 1996)
and interpreted as the result of a `superwind'
from a starburst or quasar.
However, Arp 220 has a far-infrared luminosity
100 times higher than NGC 4410A+B (\markcite{sh96}Smith $\&$ Harvey
1996), indicating 
a much higher star formation rate and/or stronger nuclear
activity than in NGC 4410A.
Better galaxies to compare with NGC 4410A are 
the
edge-on starbursts M82 and NGC 253, which
have far-infrared luminosities about five times that
of NGC 4410A+B (\markcite{r88}Rice et al. 1988).
The outflows from these two galaxies have X-ray
extents of only $\sim$10 $-$ 15 kpc and X-ray luminosities an
order of magnitude lower than that of the NGC 4410 `tail'
(\markcite{f88}Fabbiano 1988; \markcite{bst95}Bregman,
Schulman, $\&$ Tomisaka 1995; \markcite{rps97}Read,
Ponman, $\&$ Strickland 1997; \markcite{dwh98}Dahlem, Weaver,
$\&$ Heckman 1998).
This argues that the extended X-ray structure
near NGC 4410A+B is not due to a galactic wind.

It is possible that the extended X-ray emission near NGC 4410
is due to
background or foreground sources 
or coincidently superposed intracluster gas
unrelated to the bridge.
ROSAT maps of other groups and poor clusters often show
asymmetric distorted X-ray morphologies
(\markcite{d95}Doe et al. 1995; \markcite{pbe95}Pildis, Bregman,
$\&$ Evrard 1995;
\markcite{m96}Mulchaey et al. 1996; \markcite{mz98}Mulchaey $\&$
Zabludoff 1998).
An alternative hypothesis is that
the X-rays are arising from tidal gas shocked 
and heated during the interaction between the two galaxies, 
as suggested for the 
X-ray emitting gas 
in the Fornax A features by
\markcite{mf98}Mackie $\&$ Fabbiano (1998).
The spatial coincidence of
the X-ray feature and the optical bridge in NGC 4410
is excellent (Figure 4b),
suggesting an association.
Using standard X-ray cooling functions
(\markcite{mc77}McKee $\&$ Cowie 1977; 
\markcite{m87}McCray 1987)
and assuming a face-on flattened cylinder with an axial ratio between
0.1 and 1, 
the mass of X-ray emitting gas in the NGC 4410 extension
is 3 $-$ 8 $\times$ 10$^8$
M$\sun$.  This is similar to the amount inferred in the Fornax A features
(\markcite{mf98}Mackie $\&$ Fabbiano 1998), and is also similar
to the mass of HI found in the southeastern tail of NGC 4410A+B.
Perhaps in the bridge the atomic gas has been
converted to hot ionized gas, but in the southeastern 
tail the gas has remained atomic.
Further observations are needed to resolve the question
of whether the extended X-ray emission near NGC 4410A
is due to intracluster gas
or gas affiliated with the bridge.

\newpage

\subsection{A Jet-Intracluster Medium Encounter?}

\markcite{sb87}Stocke $\&$ Burns (1987) suggest that
the peculiar radio structure of NGC 4410A is 
due to motion through an intracluster medium.
The new ROSAT data provides stronger constraints
on this hypothesis while not ruling it out completely.
As noted above, NGC 4410A+B is detected in X-rays,
however, 
this emission is confined to the immediate vicinity
of the galaxies and a tail-like feature
extending to the east
(Figure 4b).
It is possible that this `tail' is associated with the
stellar bridge 
and is not truly intracluster gas (Section 4.1).
Excluding this feature, the ROSAT PSPC 3$\sigma$ upper limit to
the diffuse X-ray emission from the group is 4 $\times$ 10$^{-7}$
counts s$^{-1}$ arcsec$^{-2}$, which corresponds to 5 $\times$ 10$^{-18}$
erg s$^{-1}$ cm$^{-2}$ arcsec$^{-2}$,
assuming a 1~keV Raymond-Smith plasma and N$_{H}$ = 1.7 $\times$ 10$^{20}$
cm$^{-2}$.
This gives an average density of the intracluster medium
in the group
of $\le$6 $\times$ 10$^{-4}$ 
cm$^{-3}$, assuming
standard cooling functions (\markcite{mc77}McKee $\&$ Cowie 1977;
\markcite{m87}McCray 1987).
This is lower than the densities inferred
in many rich clusters and compact
groups (e.g., \markcite{jf84}Jones $\&$ Forman 1984; 
\markcite{pbe95}Pildis et al. 1995),
but consistent with those found in some 
poor
clusters (\markcite{d95}Doe et al. 1995),
some of which contain `head-tail' radio sources.

This upper limit on the intracluster gas density is
not inconsistent with pressure confinement of the radio
lobes by an ambient medium.
Using the method described 
by \markcite{oo87}O'Dea $\&$ Owen (1987),
we calculated the minimum pressure P$_{min}$ 
in the southeastern radio lobe of NGC 4410A+B,
assuming a spectral 
index of $\alpha$ $\sim$ 1 (F$_\nu$ $\propto$ v$^{-\alpha}$), 
lower and upper
frequency cutoffs of 10 MHz and 10 GHz respectively, a volume filling factor
of 1, and equal proton and electron energies.
Using the volume
emissivity given in 
\markcite{h86}Hummel et al. (1986),
P$_{min}$ = 
4.6 $\times$ 10$^{-13}$ dynes cm$^{-2}$ for
the diffuse emission in the lobe.
If this diffuse emission is in pressure equilibrium with 
the surrounding intracluster medium, 
$\rho$$_{ICM}$ $\sim$ 2 $\times$ 10$^{-4}$ cm$^{-3}$
(calculated as in \markcite{sb87}Stocke $\&$ Burns 1987),
in agreement with the ROSAT upper limit.

This upper limit on the amount of diffuse hot gas in the group can
be compared with the virial mass of the group.
The luminosity-weighted velocity dispersion in the NGC 4410 group
(from Table 1a) is 225 km s$^{-1}$.  Using an effective radius
of $\sim$5$'$ (140 kpc),  the virial mass is 5.1 $\times$ 10$^{12}$ M$\sun$.
This gives a M/L$_B$ ratio of $\sim$30 M$\sun$/L$\sun$ for the group, typical
of poor groups (e.g., \markcite{m96}Mulchaey et al. 1996).
From the ROSAT map, we infer an upper limit
to the mass of hot gas in this group of $\le$2 $\times$
10$^{11}$ M$\sun$, $\le$4$\%$ of the virial mass, consistent with 
the 
3$-$ 6$\%$
found for
other groups
(\markcite{pbe95}Pildis et al. 1995;
\markcite{m96}Mulchaey et al. 1996).

An intracluster medium
density of $\sim$10$^{-4}$ cm$^{-3}$ is
sufficient to produce a `head-tail' radio structure,
provided the velocity of the galaxy with
respect to the intracluster medium is high enough
or the density in the jet is low enough
(\markcite{sb87}Stocke $\&$ Burns 1987;
\markcite{v94}Venkatesan et al. 1994;
\markcite{d95}Doe et al. 1995).
For a non-relativistic jet, the radius of curvature
of the jet is R = r$_j$$\rho$$_j$v$_j$$^2$/$\rho$$_{ICM}$v$_g$$^2$
(\markcite{b81}Burns 1981; \markcite{e84}Eilek et al. 1984),
where r$_j$ is the radius of the jet,
$\rho$$_j$ and $\rho$$_{ICM}$
are the densities of the jet and the intracluster
medium respectively, v$_j$ is the velocity of
the jet, and v$_g$ is the velocity of the 
galaxy with respect to the intracluster gas.
This equation shows that a small radius of curvature can be
achieved with a sufficiently large v$_g$
or low $\rho$$_j$, even if the intracluster medium density
is low.

To test whether this condition holds in NGC 4410, we must 
estimate the parameters of the jet.
Assuming that the distorted southeastern lobe is indeed
a gradually curving jet, we estimate 
a radius of curvature R
of $\sim$1\farcm5 (40 kpc)
from the \markcite{h86}Hummel
et al. (1986) maps.
To estimate the physical conditions in this jet, we assume that
the observed radio luminosity of the lobe L$_r$ is powered
by the bulk kinetic energy in the jet, so
L$_r$ $\sim$ $\pi$$\rho$$_j$r$_j$$^2$v$_j$$^3$$\epsilon$/2,
where $\epsilon$ is the efficiency of the conversion
(\markcite{b81}Burns 1981; \markcite{e84}Eilek et al. 1984).
The radio luminosity
of the southeastern lobe 
L$_r$
is 1.3 $\times$ 10$^{41}$
erg s$^{-1}$
(\markcite{h86}Hummel et al. 1986).
Assuming that the prominent ridge extending southeast
from the nucleus 
(\markcite{h86}Hummel et al. 1986)
is the jet
gives r$_j$ $\sim$ 10$''$ (5 kpc).
Using these values and
assuming v$_j$ $\le$ 0.2c and $\epsilon$ $\le$ 0.01 
(\markcite{o85}O'Dea 1985;
\markcite{oeo93}O'Donoghue, Eilek, $\&$ Owen 1993;
\markcite{bc93}Borne $\&$ Colina 1993),
we find that the internal ram pressure
in the jet is $\rho$$_j$v$_j$$^2$ $\ge$ 4 $\times$ 10$^{-12}$ dynes cm$^{-2}$.
For comparison,
using the assumptions given above and the 
\markcite{h86}Hummel et al. (1986)
volume emissivity,
the minimum pressure P$_{min}$ 
in the more northern of the radio knots in the southeastern lobe
is 1.3 $\times$ 10$^{-12}$ dynes cm$^{-2}$,  
consistent with this estimate.

Using our estimate of the internal ram pressure in the 
NGC 4410 jet, we find that 
the velocity of the galaxy relative to
the intracluster medium v$_g$ must be greater than 220 km~s$^{-1}$
to have caused
the observed bending of the radio lobe.
A more moderate jet velocity of $\sim$10$^4$ km s$^{-1}$ (e.g., 
\markcite{bc93}Borne $\&$ Colina 1993) gives a lower limit
of v$_g$ $\ge$ 660 km s$^{-1}$.
The radial velocity of NGC 4410A, 7440 km s$^{-1}$, is close to
the luminosity-weighted mean velocity of the group, 7355 km s$^{-1}$.
Therefore, 
if the distortion of the lobe was caused
by ram pressure bending by an intracluster medium at rest
with respect to the group, 
much of this motion must be in the plane of the sky. 
Optically, NGC 4410A is the most luminous galaxy in the group, 
providing 1/4 of its total blue luminosity.
It has 
an absolute blue magnitude of $-$21.1, approaching that of M87.
This argues against the idea that NGC 4410A has a large 
motion relative
to the group as a whole.
Therefore, if the group is virialized,
then the distortion of the radio lobe 
may not have been caused by an interaction with an intracluster medium.

\subsection{A Jet-Interstellar Medium Encounter?}

The HI tail to the southeast of NGC 4410A+B, overlapping with
the bright southeastern radio lobe, suggests an intriguing 
alternative explanation for
the peculiar radio morphology in this system: perhaps it
is not
due to an interaction with an intracluster medium, as in 
classical models of head-tail systems,
but instead is due to an interaction with the interstellar medium
in the system.  
Interactions between jets and interstellar gas disturbed
by a gravitational encounter or merger between galaxies
have been suggested for a number of systems,
including 4C 29.30 (\markcite{vB86}van Breugel et al. 1986)
and Centaurus
A (\markcite{dY81}de Young 1981).
Although it is possible that 
the southeastern tidal tail and the radio lobe of 4410A+B are merely superposed
structures, extending out from the pair of galaxies
in different directions, 
the apparent abrupt bend 
in the radio lobe 
(\markcite{h86}Hummel et al.
1986) suggests that they are interacting,
as does
the discontinuity in the HI velocity structure of this tail
(Figure 3).

It is uncertain at present if there is sufficient
interstellar matter in the southeastern tail
to account for the bending of the lobe,
because of the lack of information
about the properties of the jet and the timescale of
the impact.
A gas cloud of mass $M_c$ impacted by a jet has an acceleration
a$_c$ given by $M_ca_c = {\rho}_jv_j^2{\pi}r_j^2$
(\markcite{e84}Eilek et al. 1984).
Using the relationship provided above for the jet parameters
in terms of the radio luminosity L$_r$ gives
$M_c \sim 2L_rt_c/{\epsilon}v_jv_c$, assuming the cloud
has been accelerated to a velocity v$_c$ over a period t$_c$.
Assuming v$_c$ $\sim$ 240 km s$^{-1}$
(from Figure 3; the maximum radial velocity difference between the HI
in the tail and the nucleus) and 
$\epsilon$ $\sim$ 0.01,
$M_c$ $\sim$ 3 $\times$ 10$^6$ (60000/$v_j$ km s$^{-1}$)/($t_c$/10$^6$
years)
M$\sun$.
For comparison, the mass of HI in this tail is 3 $\times$ 10$^8$ M$\sun$
(Section 3.1.2).  Therefore, if the timescale is short and/or the jet velocity
is high, there is sufficient interstellar gas to bend the jet.
The timescale t$_c$ is uncertain, but
it must be less than the age of the radio source.
Therefore t$_c$ $\le$ D$_{northern~lobe}$/v$_j$, where the D$_{northern~
lobe}$ is the spatial extent of the northern lobe (180 kpc).
As noted above, the jet velocity v$_j$ is uncertain, but is probably
less than 0.2c
(e.g., \markcite{o85}O'Dea 1985;
\markcite{oeo93}O'Donoghue, Eilek, $\&$ Owen 1993;
\markcite{bc93}Borne $\&$ Colina 1993).
If v$_j$ is close to this upper limit, then
t$_c$ $\le$ 3 $\times$ 10$^6$ years and
the required mass M$_c$ $\le$ 9 $\times$ 10$^6$ M$\sun$,
less than the amount of interstellar gas in the tail.
If v$_j$ is small and the timescale of impact is long,
however, 
the gas in the tail may not be sufficient to bend the
radio
structure. 

A related question is whether the gas in the tail is able to
bend the jet to the observed angle
($\sim$70$^{\circ}$; \markcite{h86}Hummel et al. 1986). 
This angle is larger than the bends in some radio structures
where jet-cloud interactions are thought to have occurred
(\markcite{vB85a}van Breugel et al. 1985a, \markcite{vB86}1986;
\markcite{sbc85}Stocke et al. 1985).
However, there are 
a few other examples of radio lobes bent at large angles
(\markcite{sbc85}Stocke et al. 1985;
\markcite{bc93}Borne $\&$ Colina 1993), so NGC 4410 is not unique.
There have been a number of theoretical studies
of jet bending in the literature (e.g.,
\markcite{v84}Valtaoja 1984; \markcite{fh84}Fiedler $\&$ Henriksen 1984;
\markcite{wg85}William $\&$ Gull 1985;
\markcite{sbc85}Stocke et al.
1985;  
\markcite{lb86}Lonsdale $\&$ Barthel 1986; 
\markcite{bc93}Borne $\&$ Colina 1993).
Of these models, the one that may be the most relevant for NGC 4410
is the \markcite{bc93}Borne $\&$ Colina (1993) simulation
of the 70$^{\circ}$ bend in the eastern 3C 278 radio lobe from
ram pressure due to the interstellar medium in a companion, caused by
the companion's motion relative to the radio jet.
Using estimated
orbital parameters of the host
galaxy NGC 4782 and its companion NGC 4783,
\markcite{bc93}Borne $\&$ Colina (1993) successfully
reproduced 
the observed bending of the 3C 278 jet.
Their model demonstrates that ram pressure due to the interstellar matter
in a companion moving past a radio galaxy is indeed capable of
bending a jet to large angles.

\subsection{A Group-Group Merger?}

If the group is dynamically young,
the argument given in Section 4.2 
against ram pressure bending by
the intracluster medium does not hold
(e.g., \markcite{v94}Venkatesan et al. 1994; \markcite{d95}Doe
et al. 1995).
It is possible that there is a large-scale motion of the group
relative to an intracluster medium, perhaps
due to the merging of two groups.
\markcite{b98}Bliton et al. (1998) compared the 
radial velocities of a sample of 23 head-tail galaxies in clusters with the 
cluster mean velocities and dispersions, 
and concluded
that the relative velocities are too small to account for the distortion
of the radio lobes.
They suggested that these systems are undergoing cluster-cluster mergers,
which are driving bulk flows of the intracluster medium.
Such a process may be occurring in the NGC 4410 group;
two of the galaxies in the group, NGC 4410D and VCC 822, 
have radial velocities that differ from the mean by 500 km s$^{-1}$.
If the extended X-ray emission noted earlier is indeed
intracluster gas rather than material associated with the
bridge, 
then the intracluster gas is quite inhomogeneous,
supporting the idea of a group-group merger.
To test this possibility, 
higher sensitivity X-ray data of the NGC 4410 group would be useful.

It is possible that the on-going interaction between
NGC 4410A and NGC 4410B has been driven by the merger of
two groups of galaxies, and the disturbance of the radio structure
was caused by an interaction with both the interstellar
and the intracluster medium, both having been perturbed by
the encounter.
The radio morphology of NGC 4410 resembles that of 
the peculiar system 
3C 442 
in some respects.
\markcite{co91}Comins $\&$ Owen (1991) suggest that
the 
distorted radio lobes
and prominent ridges 
found in 3C 442 were created during the merger of two clusters.
The host galaxy of 3C 442 is currently merging with a companion galaxy;
if this galaxy-galaxy merger was driven by a cluster-cluster merger,
in addition to perturbed interstellar matter, there may be
disturbed intracluster matter which could be disrupting the radio
lobes.
Based upon the observation of a relatively constant 6 to 20 cm spectral
index of $\alpha$ $\sim$ 1.1 
along the 3C 442 lobe,
\markcite{co91}Comins $\&$ Owen (1991) suggest that 
electrons in the 3C 442 lobes
may have been re-accelerated 
by turbulence from the interaction.
The southeast lobe of NGC 4410 also has a relatively
constant spectral index of $\alpha$ $\sim$ 1.0
(\markcite{h86}Hummel
et al. 1986),
arguing for
re-acceleration in that structure as well.
Whether this re-acceleration and the observed distortion
of the radio lobe were caused by an encounter with the
interstellar or intracluster matter is as yet uncertain.
However, the existence of a considerable amount of
interstellar matter in the NGC 4410 system, along with the non-detection
of diffuse intracluster medium in the group,
argues that
interstellar gas played a major role in the shaping of the radio structure
in this system.

\section{CONCLUSIONS}

Using the VLA and the NRAO 12m telescope, we mapped
the radio continuum emission, the 21 cm HI line,
and the 2.6mm CO line in the peculiar interacting galaxy
pair NGC 4410A+B.
We 
discovered a low surface brightness radio
lobe extending 6\farcm2 (180 kpc) to the northwest of the
pair,
opposite the bright southeastern lobe seen in
previous observations.
The existence of this second lobe confirms
that NGC 4410A is a double-lobed radio galaxy, albeit
with a severely distorted structure.

We have detected both HI and CO in 
this system, yielding M$_{HI}$ $\sim$ 10$^9$ M$_{\sun}$
and M$_{H_2}$ $\sim$ 4 $\times$ 10$^9$ M$_{\sun}$.
The molecular gas is concentrated in
the radio galaxy NGC 4410A while the HI is more extended.
There is a 
3 $\times$ 10$^8$ M$\sun$ 
HI tail 
extending 
1\farcm7 (50 kpc) 
to the southeast
of the pair,
coincident with an optical tail.
This HI tail overlaps part of the southeastern
radio lobe.
Comparison with a recent ROSAT PSPC X-ray map of this system shows
an anti-coincidence of the HI with an X-ray feature
that extends due east from NGC 4410A.  This X-ray
structure
contains 3$-$8 $\times$ 10$^8$ M$_{\sun}$
of hot gas
and appears to be aligned with
a stellar bridge connecting the pair to a third galaxy.
It is not clear at present whether the extended X-ray emitting gas
is intracluster material or shocked gas associated with the
bridge.

We suggest that the distortion of the radio structure
was caused by ram pressure 
from the interstellar medium in the system,
due to the motion of the interstellar matter relative to the radio source
during
the gravitational interaction of NGC 4410A with
NGC 4410B.
An alternative possibility is that the peculiar radio 
structure may
be due to motion through an intragroup
medium, however, the small radial velocity of NGC 4410A relative
to the group and the lack of diffuse X-ray emission in the
group makes 
this scenario somewhat less likely, unless the group is not
virialized or is merging with another group.

In addition to NGC 4410A+B, we detected
HI in 
six other
galaxies in the NGC 4410 group.
The total HI in the group is 1.4 $\times$ 10$^{10}$ M$\sun$, 80$\%$ of which
arises from four galaxies in the outer group.
Three of these galaxies (VCC 831,
VCC 822, and VCC 847) are spirals with M$_{HI}$/L$_B$ ratios
typical of field galaxies,
while FGC 170A
appears to be a gas-rich dwarf galaxy.
In the inner group, 
the SBa galaxy
NGC 4410D (VCC 934)
was detected in both HI and CO
and has a 
2 $\times$ 10$^8$ M$_{\sun}$ 
HI tail 
extending
1$'$ (28 kpc) 
towards the nearby disk galaxy NGC 4410F.
NGC 4410F was also seen in HI
but not in CO.
The galaxies in the inner group appear to be somewhat deficient
in HI compared to their blue luminosities,
suggesting phase changes driven by galaxy-galaxy or galaxy-intracluster
medium encounters.

\vskip 0.2in

We would like to thank Liese van Zee, Michael
Rupen, and Jason Wurnig for their help in
planning for and reducing the
VLA observations.
We also appreciate the help of the telescope
operators and staff of
the NRAO 12m telescope in obtaining the CO data.
We thank the referee Jacqueline van Gorkom, Frazer Owen, Curt Struck, Daniel Tsch\"oke,
and Bev Wills for helpful comments on this paper,
and Daniel Tsch\"oke and Megan Donahue
for providing us with copies of their X-ray and optical maps.
This research has made use of the NASA/IPAC Extragalactic
Database (NED) which is operated by the Jet Propulsion
Laboratory under contract with NASA.
We are pleased to acknowledge partial funding for
this project from a NASA grant administrated
by the American Astronomical Society.  Partial support
was also provided from NASA grant AR-08374.01-97A
from the Space Telescope Science Institute,
which is operated by AURA, Inc., under NASA contract NAS 5-26555.

\vfill
\eject

{\bf Captions}

\figcaption[BSmith_fig1.ps]{a) A 500 second archival
Hubble Space Telescope
(HST) WFPC2 image of NGC 4410, taken with the broadband red F606W filter.
These data were originally obtained as part of the 
Malkan et al. (1998) HST survey of active galaxies,
and have previously been presented by 
Tsch\"oke et al. (1999).
b) A 1\farcm1 $\times$ 0\farcm9 field of view 
narrowband red continuum image of NGC 4410A+B, obtained with
the Kitt Peak 2.1m telescope (Donahue et al. 2000).
NGC 4410A is the more western galaxy.
c) The inner 
6\farcm4 $\times$ 4\farcm9
of the NGC 4410 group, as seen in a deep R band image 
from the Southeastern Association for Research in Astronomy
(SARA) telescope
(from Donahue et al. 2000).
The five major members are labeled.
d) The Digitized Sky Survey image of the NGC 4410 group.
The 23 known galaxies in the field are marked,
as well as four anonymous galaxies (see Tables 1a $-$ c).
Twelve of these galaxies are part of the NGC 4410 group,
two (NGC 4411 and NGC 4411B) belong to the foreground Virgo Cluster,
and four to a background group.  The rest of the galaxies
have no published redshifts.
}

\figcaption[BSmith_fig2.ps]{
Our 
naturally-weighted 20 cm continuum map of the NGC 4410 field (contours),
superposed on the Digitized Sky Survey map (greyscale).
The first contour is 0.75 mJy beam$^{-1}$ and the contour
intervals increase by multiples of $\sqrt2$. 
The beamsize is 59\farcs38 $\times$ 54\farcs67.
For clarity, this map has not been corrected for primary beam attenuation.
The noise level in the inner portion of the map is 0.17 mJy beam$^{-1}$.
}

\figcaption[BSmith_fig3.ps]
{The naturally-weighted
HI channel maps of the inner portion of the NGC 4410
group (contours),
superposed on the optical DSS map (greyscale). 
The contour levels are $-$2$\sigma$, 2$\sigma$,
4$\sigma$, 6$\sigma$, 8$\sigma$, and 10$\sigma$,
where 1$\sigma$ = 0.21 mJy beam$^{-1}$.
The HI beamsize 
in this map
is 59\farcs38 $\times$ 54\farcs67.
}

\figcaption[BSmith_fig4.ps]
{a)
The naturally-weighted HI intensity map of the inner
portion of the NGC 4410 group (contours),
superposed on the R band image from
Donahue et al. (2000) (greyscale).
The first contour is 2 $\times$ 10$^{19}$ cm$^{-2}$;
the contour interval is 10$^{19}$ cm$^{-2}$.
The HI beamsize is 59\farcs38 $\times$ 54\farcs67.
b) 
The HI intensity map (blue contours),
the radio continuum map (white contours), and the ROSAT
PSPC X-ray map (yellow contours), all
superposed on the 
R band image (greyscale).
The first radio continuum contour is 0.75 mJy beam$^{-1}$ and the contour
intervals increase by multiples of $\sqrt2$. 
The contour levels on the X-ray maps are 3$\sigma$, 4$\sigma$, 7$\sigma$,
12$\sigma$, 19$\sigma$, 28$\sigma$, 39$\sigma$, and 52$\sigma$
above the background, where the background is 1.4 $\times$ 10$^{-2}$ counts
arcsec$^{-2}$
and the rms noise is 3 $\times$ 10$^{-3}$ counts arcsec$^{-2}$.
The spatial resolution of the PSPC map is 25$''$, and 
the total integration time with ROSAT was 23.4 ksec.
}

\figcaption[BSmith_fig5.ps]
{A larger field of view display
of the naturally-weighted HI intensity
map (contours), showing the entire NGC 4410 group.
This image is superposed on the optical DSS image (greyscale).
The first contour is 3 $\times$ 10$^{19}$ cm$^{-2}$ 
and the contour interval is 6 $\times$ 10$^{19}$ cm$^{-2}$.
The increased noise at the edges of the field is due
to the primary beam correction.
The beamsize is 59\farcs38 $\times$ 54\farcs67.
}

\figcaption[BSmith_fig6.ps]
{a-b) The total 21 cm HI spectra for the detected galaxies
in the NGC 4410 group.  These have been corrected for primary
beam attenuation.}

\figcaption[BSmith_fig7.ps]
{The CO spectra for the observed galaxies in the NGC 4410 group.
For displaying purposes, these spectra 
have been smoothed by a 36 km s$^{-1}$ boxcar and then resampled
at 21 km s$^{-1}$ spacing.}

\figcaption[BSmith_fig8.ps]
{a-d) The naturally-weighted
HI MOM1 maps (contours) for the four detected galaxies in the outer
part of the NGC 4410 group,
superposed on the optical DSS images (greyscale).  
The contour intervals are 10 km s$^{-1}$.  Several
contours are 
labeled, in units of km s$^{-1}$.}

\vfill
\eject

\begin{table}
   {\bf Table 1a}\\
   Galaxies in the inner 30$'$ Diameter of the NGC 4410 Group$^a$ \\ [12pt]
   \begin{tabular}{crrrrrrrrcccclclcrcccc} \hline
       \multicolumn{1}{c}{Name}&
\multicolumn{6}{c}{Position}
&\multicolumn{1}{c}{Type}
&\multicolumn{1}{c}{m$_B$}
&\multicolumn{1}{c}{Velocity}\\
       \multicolumn{1}{c}{}&
\multicolumn{3}{c}{R.A. (1950)}&
\multicolumn{3}{c}{Dec. (1950)}&
\multicolumn{1}{c}{}& 
\multicolumn{1}{c}{}& 
\multicolumn{1}{c}{(km/s)}\\
\hline
NGC 4410A$^b$&12&23&55.7&9&17&47&Sab Pec?&13.8&7440$^c$&&\\
NGC 4410B$^b$&12&23&57.2&9&17&46&S0? Pec&15.0&7500$^c$&\\
NGC 4410C (IC 790)&12&24&2.9&9&18&43&S0?&15.5&7525\\
NGC 4410D (VCC 934)&12&24&11.7&9&19&31&SBa(s)&14.8&6938
\\
NGC 4410E (LEDA 094212)&12&23&59.1&
9&23&59&S$^d$&16.0&7355\\
NGC 4410F$^b$&12&24&8.2&9&17&30&&16$^d$&7206$^e$\\
NGC 4410H (VCC 961)&12&24&26.6&9&14&7&
SB0&15.0&7342\\
VCC 822&12&23&8.0&9&11&28&SBc&15.0&6898\\
VCC 831&12&23&11.0&9&18&12&Sc&15.0&7507\\
VCC 847&12&23&19.0&9&12&2&SBc&14.9&7556\\
FGC 170A&12&24&34.0&9&29&54&S$^d$&17$^d$&7620$^f$\\
NGC 4410K (1224+0919)&12&24&56.7&9&19&11&&15$^d$&7458\\

\tablenotetext{ }{}
\tablenotetext{a}{All
information from the NASA/IPAC Extragalactic
Database (NED), unless otherwise noted.
The nomenclature used for NGC 4410A$-$H, K
is from \markcite{sb87}Stocke $\&$ Burns (1987).
The velocities are optical heliocentric values.
}
\tablenotetext{b}{Position
measured from the Digitized Sky Survey (DSS) image.
}
\tablenotetext{c}{Optical velocity from \markcite{d99}Donahue et al. (2000).
}
\tablenotetext{d}{Estimated
from the Digitized Sky Survey (DSS) image.
}
\tablenotetext{e}{Optical
velocity from Stocke $\&$ Burns (1987), converted from v$_{LSR}$ to v$_{helio}$
as in the RC3.
}
\tablenotetext{f}{HI
velocity from this work.
}
\end{tabular}
\end{table}

\newpage

\begin{table}
   {\bf Table 1b}\\
   Foreground and Background Objects in the NGC 4410 Field$^a$ \\ [12pt]
   \begin{tabular}{crrrrrrrrcccclclcrcccc} \hline
       \multicolumn{1}{c}{Name}&
\multicolumn{6}{c}{Position}
&\multicolumn{1}{c}{Type}
&\multicolumn{1}{c}{m$_B$}
&\multicolumn{1}{c}{Velocity}\\
       \multicolumn{1}{c}{}&
\multicolumn{3}{c}{R.A. (1950)}&
\multicolumn{3}{c}{Dec. (1950)}&
\multicolumn{1}{c}{}& 
\multicolumn{1}{c}{}& 
\multicolumn{1}{c}{(km/s)}\\
\hline
NGC 4411$^b$&12&23&56.6&9&8&53&SB(rs)c&13.4&1281\\
NGC 4411B$^b$&12&24&14.7&9&9&41&SAB(s)cd&12.9&1270\\
LEDA 094219&12&24&53.0&9&13&38&S$^c$&17.3&
26372\\
NGC 4410I (Abell 1541)&12&24&15.6&9&14&17&&16$^c$&25828\\
NGC 4410G$^d$ (LEDA 094214)&12&24&25.0&9&17&41&&17.2&27462\\
NGC 4410J$^d$&12&24&31.3&9&22&19&S$^c$&16$^c$&25066$^e$\\

\tablenotetext{}{}
\tablenotetext{a}{All
information from the NASA Extragalactic
Database (NED), unless otherwise noted.
The nomenclature used for NGC 4410G, I, and J
is from \markcite{sb87}Stocke $\&$ Burns (1987).
}
\tablenotetext{b}{Member of the Virgo Cluster (\markcite{bst85}Binggeli et al. 1985).
}
\tablenotetext{c}{Estimated
from the Digitized Sky Survey (DSS) image.
}
\tablenotetext{d}{
Position measured from the DSS.
}
\tablenotetext{e}{Optical
velocity from Stocke $\&$ Burns (1987), converted from v$_{LSR}$ to v$_{helio}$
as in the RC3.
}
\end{tabular}
\end{table}

\newpage

\begin{table}
   {\bf Table 1c}\\
   Galaxies in the NGC 4410 Field with Unknown Velocities$^a$ \\ [12pt]
   \begin{tabular}{crrrrrrrrcccclclcrcccc} \hline
       \multicolumn{1}{c}{Name}&
\multicolumn{6}{c}{Position}
&\multicolumn{1}{c}{Type}
&\multicolumn{1}{c}{m$_B$}\\
       \multicolumn{1}{c}{}&
\multicolumn{3}{c}{R.A. (1950)}&
\multicolumn{3}{c}{Dec. (1950)}&
\multicolumn{1}{c}{}& 
\multicolumn{1}{c}{}\\
\hline
VCC 869$^b$&12&23&33.3&9&14&39&ImV or dE0&15.0\\
VCC 914$^b$&12&24&1.2&9&16&12&dE,N&19.0\\
VCC 933$^b$&12&24&11.4&9&6&0&dE2,N&16.6&\\
VCC 902$^b$&12&23&55.2&9&4&24&dE0&18.3\\
VCC 976$^b$&12&24&39.0&9&6&54&dE4&18.0\\
ANON 1$^c$&12&22&57.6&9&20&55.6&&16$^d$\\
ANON 2$^c$&12&23&12.6&9&16&27&&16.5$^d$\\
ANON 3$^c$&12&24&0.3&9&19&36&&17$^d$\\
ANON 4$^c$&12&24&36.8&9&27&38&&16$^d$\\

\tablenotetext{ }{}
\tablenotetext{a}{All information from the NASA Extragalactic
Database (NED), unless otherwise noted.
}
\tablenotetext{b}{Based on optical morphology, \markcite{bst85}Binggeli et al.
(1985) conclude this is probably a member of the Virgo Cluster.}
\tablenotetext{c}{Position
measured from the Digitized Sky Survey (DSS) image.
}
\tablenotetext{d}{Estimated
from the Digitized Sky Survey (DSS) image.
}
\end{tabular}
\end{table}

\begin{table}
   {\bf Table 2}\\
   VLA 21 cm HI Results for the NGC 4410 Group \\ [12pt]
   \begin{tabular}{ccccclclcrcccc} \hline
       \multicolumn{1}{c}{Name}
&\multicolumn{1}{c}{HI Line}
&\multicolumn{1}{c}{Central}
&\multicolumn{1}{c}{$\Delta$V}\\
       \multicolumn{1}{c}{}&
\multicolumn{1}{c}{Flux}& 
\multicolumn{1}{c}{Velocity}& 
\multicolumn{1}{c}{(FWHM)}\\
       \multicolumn{1}{c}{}&
\multicolumn{1}{c}{(Jy km s$^{-1}$)}&
\multicolumn{1}{c}{(km s$^{-1}$)}& 
\multicolumn{1}{c}{(km s$^{-1}$)}&\\ 
\hline
NGC 4410A+B&0.55 $\pm$ 0.16&7350&260\\
NGC 4410C&$\le$0.08$^a$\\
NGC 4410D&0.22 $\pm$ 0.04&6960,7380&90,90\\
NGC 4410E&$\le$0.10$^a$\\
NGC 4410F&0.17 $\pm$ 0.05&7180&90\\
NGC 4410H&$\le$0.09$^a$\\
VCC 822&1.63 $\pm$ 0.13&6900$^c$&200$^c$\\
VCC 831&0.74 $\pm$ 0.11&7430&220\\
VCC 847&1.14 $\pm$ 0.07&7550$^b$&250$^b$\\
FGC 170A&1.21 $\pm$ 0.13&7620$^b$&160$^b$\\
1224+0919&$\le$0.16$^a$\\
\\
\tablenotetext{ }{}
\tablenotetext{a}{Assuming a 
line width of $\le$600 km s$^{-1}$ and
a $\le$30$''$ (14 kpc) diameter HI disk.}
\tablenotetext{b}{Near the edge of the observed
bandpass.  Some high velocity gas may have been missed.}
\tablenotetext{c}{Near the edge of the observed
bandpass.  Some low velocity gas may have been missed.}
   \end{tabular}
\end{table}

\begin{table}
   {\bf Table 3}\\
   NRAO 12m CO (1-0) Results for the NGC 4410 System \\ [12pt]
   \begin{tabular}{crrrrrrrrcccclclcrcccc} \hline
       \multicolumn{1}{c}{Name}
&\multicolumn{1}{c}{T$_R$$^*$ (rms)}
&\multicolumn{1}{c}{$I_{CO}$$^a$}
&\multicolumn{1}{c}{Velocity}
&\multicolumn{1}{c}{$\Delta$V$^b$}\\
       \multicolumn{1}{c}{}&
\multicolumn{1}{c}{(mK)}& 
\multicolumn{1}{c}{(K km s$^{-1}$)}& 
\multicolumn{1}{c}{(km s$^{-1}$)}& 
\multicolumn{1}{c}{(km s$^{-1}$)}&\\
\hline
NGC 4410A&1.5&1.12 $\pm$ 0.09&7140,7300,7565&600\\
NGC 4410B&1.6&$\le$0.30$^c$\\
NGC 4410A+B Tail$^d$&1.5&$\le$0.18$^e$\\
NGC 4410C&2.8&$\le$0.47$^f$\\
NGC 4410D&2.1&0.24 $\pm$ 0.08&6960&150\\
NGC 4410F&2.1&$\le$0.14$^g$\\
\\

\end{tabular}
\tablenotetext{ }{}
\tablenotetext{a}{Statistical uncertainties only.  Upper
limits are 3$\sigma$.}
\tablenotetext{b}{Full width half maximum line widths (FWHM).}
\tablenotetext{c}{Assuming a line width
of $\le$800 km s$^{-1}$.}
\tablenotetext{d}{
Centered at 12$^{\rm h}$ 23$^{\rm m}$ 58.3$^{\rm s}$ 9$^{\circ}$
17$'$ 11\farcs5 (1950).
}
\tablenotetext{e}{Using the HI line width of 300
km s$^{-1}$ (Figure 3).}
\tablenotetext{f}{Assuming a line width
of $\le$600 km s$^{-1}$.}
\tablenotetext{g}{Using the FWHM HI line width
of 90 km s$^{-1}$ (Table 2)}.

\end{table}

\vfill
\eject

\begin{table}
   {\bf Table 4}\\
Global Parameters for the Galaxies in the NGC 4410 Group
\\ [12pt]
   \begin{tabular}{crrrrrrrrcccclclcrcccc} \hline
       \multicolumn{1}{c}{Name}&
\multicolumn{1}{c}{M$_{HI}$}
&\multicolumn{1}{c}{M$_{H_2}$$^a$}
&\multicolumn{1}{c}{L$_B$$^b$}
&\multicolumn{1}{c}{M$_{HI}$/L$_B$}
&\multicolumn{1}{c}{M$_{H_2}$/M$_{HI}$}
\\
       \multicolumn{1}{c}{}&
\multicolumn{1}{c}{(M$_{\sun}$)}&
\multicolumn{1}{c}{(M$_{\sun}$)}&
\multicolumn{1}{c}{(L$_{\sun}$)}&
\multicolumn{1}{c}{(M$_{\sun}$/L$_{\sun}$)}&\\
\hline
NGC 4410A+B&1.3 $\times$ 10$^9$&3.9 $\times$ 10$^{9}$&
5.7 $\times$ 10$^{10}$&0.022&3.0\\
NGC 4410C&$\le$1.8 $\times$ 10$^8$
&$\le$1.7 $\times$ 10$^{9}$
&9.0 $\times$ 10$^9$&$\le$0.020\\
NGC 4410D&5.0 $\times$ 10$^8$&8.4 $\times$ 10$^8$&1.7 $\times$ 10$^{10}$&0.029&
1.7\\
NGC 4410E&$\le$2.2 $\times$ 10$^8$&&5.6 $\times$ 10$^9$&$\le$0.039\\
NGC 4410F&3.8 $\times$ 10$^8$&$\le$5.0 $\times$ 10$^8$&5.6 $\times$ 10$^9$&
0.068&$\le$1.3\\
NGC 4410H&$\le$2.0 $\times$ 10$^8$&&1.4 $\times$ 10$^{10}$&$\le$0.014\\
VCC 822&3.6 $\times$ 10$^9$&&1.4 $\times$ 10$^{10}$&0.26\\
VCC 831&1.7 $\times$ 10$^9$&&1.4 $\times$ 10$^{10}$&0.12\\
VCC 847&2.5 $\times$ 10$^9$&&1.6 $\times$ 10$^{10}$&0.16\\
FGC 170A&2.6 $\times$ 10$^9$&&2.2 $\times$ 10$^9$&1.20\\
1224+0919&$\le$3.6 $\times$ 10$^8$&&1.4 $\times$ 10$^{10}$&$\le$0.026\\
\\
\end{tabular}
\tablenotetext{ }{}
\tablenotetext{a}{Calculated
assuming the standard Galactic I$_{CO}$/N$_{H_2}$ ratio 
(M$_{H_2}$ = 1.1 $\times$ 10$^4$ D$^2$ $\int$S$_V$dV, where
D is the distance in Mpc; \markcite{b86}Bloemen et al. 1986), H$_o$
= 75 km s$^{-1}$ Mpc$^{-1}$, and assuming a point source
($\eta$$_c$ = 1) and
34 Jy K$^{-1}$ for the 12m
telescope.
}
\tablenotetext{b}{Assuming M$_{B_{\sun}}$ = 5.48.}
\end{table}

\end{document}